\documentclass[]{spie}  
\usepackage[]{graphicx}
\usepackage[]{amsmath}
\usepackage{amssymb}
\usepackage{epstopdf}

\newcommand{\pb}{\protect\textsc{polarbear}}
\newcommand{\Pb}{\protect\textsc{Polarbear}}
\newcommand{\LB}{LiteBIRD\ }

\title{Multi-chroic dual-polarization bolometric detectors for studies of the Cosmic Microwave Background} 

\author{Aritoki Suzuki\supit{a}, Kam Arnold\supit{a}, Jennifer Edwards\supit{b}, Greg Engargiola\supit{c}, Adnan Ghribi\supit{a}, William Holzapfel\supit{a}, Adrian T. Lee\supit{a,d}, Xiao Fan Meng\supit{e}, Michael J. Myers\supit{a}, Roger O'Brient\supit{f}, Erin Quealy\supit{a}, Gabriel Rebeiz\supit{b}, Paul Richards\supit{a}, Darin Rosen\supit{a}, Praween Siritanasak \supit{g}
\skiplinehalf
\supit{a}University of California, Berkeley, Physics, Berkeley, CA 94720 USA; \\
\supit{b}University of California, San Diego, Electrical and Computer Engineering, La Jolla, CA 92092 USA; \\
\supit{c}Lawrence Berkeley National Laboratory, Berkeley, CA 94720 USA; \\
\supit{d}University of California Berkeley, Radio Astronomy, Berkeley, CA 94720 USA; \\
\supit{e}University of California Berkeley, Electrical Engineering, Berkeley, CA 94720 USA; \\
\supit{f}California Institute of Technology, Physics, Pasadena, CA 91125 USA; \\
\supit{g}University of California, San Diego, Physics, La Jolla, CA 92092 USA
}

\authorinfo{Author Information: Aritoki Suzuki, e-mail: asuzuki@berkeley.edu, telephone: +1(510)642-7801}

\begin{document} 
\maketitle 

\begin{abstract}
We are developing multi-chroic antenna-coupled TES detectors for CMB polarimetry. Multi-chroic detectors increase the mapping speed per focal plane area and provide greater discrimination of polarized galactic foregrounds with no increase in weight or cryogenic cost. In each pixel, a silicon lens-coupled dual polarized sinuous antenna collects light over a two-octave frequency band. The antenna couples the broadband millimeter wave signal into microstrip transmission lines, and on-chip filter banks split the broadband signal into several frequency bands. Separate TES bolometers detect the power in each frequency band and linear polarization. We will describe the design and performance of these devices and present optical data taken with prototype pixels. Our measurements show beams with percent level ellipticity, percent level cross-polarization leakage, and partitioned bands using banks of 2, 3, and 7 filters. We will also describe the development of broadband anti-reflection coatings for the high dielectric constant lens. The broadband anti-reflection coating has approximately 100\% bandwidth and no detectable loss at cryogenic temperature. Finally, we will describe an upgrade for the \Pb\ CMB experiment and installation for the \LB CMB satellite experiment both of which have focal planes with kilo-pixel of these detectors to achieve unprecedented mapping speed.
\end{abstract}


\section{INTRODUCTION}\label{sec:intro}
Characterization of the Cosmic Microwave Background (CMB) B-mode polarization signal will test models of inflationary cosmology, as well as constrain the sum of the neutrino masses and other cosmological parameters. The low intensity of the B-mode signal combined with the need to remove polarized galactic foregrounds requires an extremely sensitive millimeter receiver and effective methods of foreground removal. Current bolometric detector technology is reaching the limit set by the CMB photon noise. Thus, we need to increase the optical throughput to increase an experiment's sensitivity. To increase the throughput without increasing the focal plane size, we can increase the frequency coverage of each pixel. Increased frequency coverage per pixel has additional advantage that we can split the signal into frequency bands to obtain spectral information. The detection of multiple frequency bands allows for removal of the polarized foreground emission from synchrotron radiation and thermal dust emission, by utilizing its spectral dependence. Traditionally, spectral information has been captured with a multi-chroic \textit{focal plane} consisting of a heterogeneous mix of single-color pixels. To maximize the efficiency of the focal plane area, we developed a multi-chroic \textit{pixel}. This increases the number of pixels per frequency, at no extra cost per focal plane area, weight and cryogenic cost. For earlier development of this work, please refer to O'Brient {\it et al.} \cite{OBrientSPIE}.

In section \ref{sec:pixel}, we discuss the design and fabrication of the multi-chroic pixel. In section \ref{sec:test}, we describe the test setup. In section \ref{sec:result}, results are presented. We show our development in broadband anti-reflection (AR) coating in section \ref{sec:ar}. Finally, we conclude in section \ref{sec:conclusion} with fabrication of arrays of multi-chroic detectors and plans for implementing these multi-chroic detector arrays in next-generation CMB experiments.

\section{PIXEL DESIGN}\label{sec:pixel}
\begin{figure}
\begin{center}
\begin{tabular}{c}
\includegraphics[height=4cm,keepaspectratio]{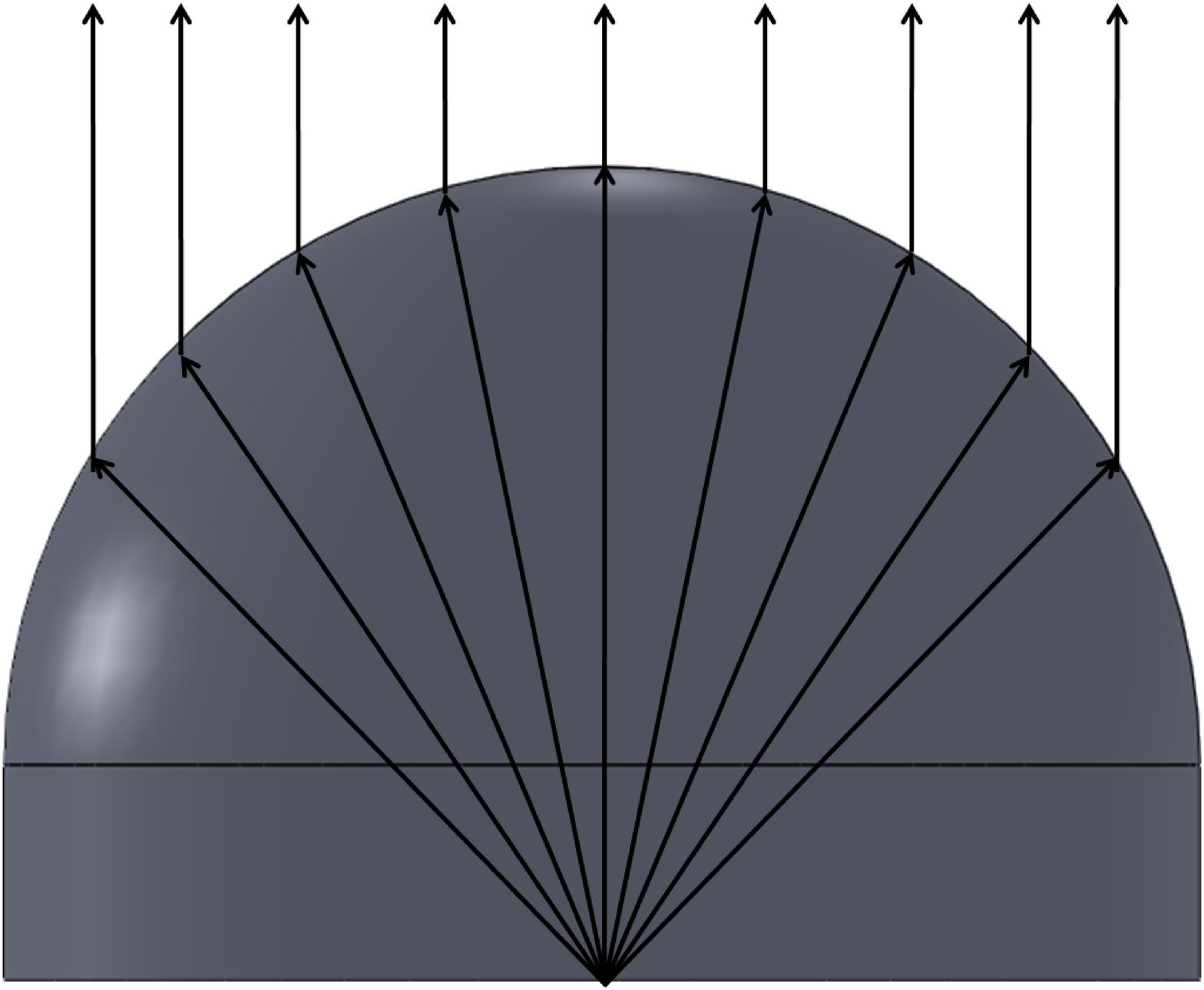}
\includegraphics[height=4cm,keepaspectratio]{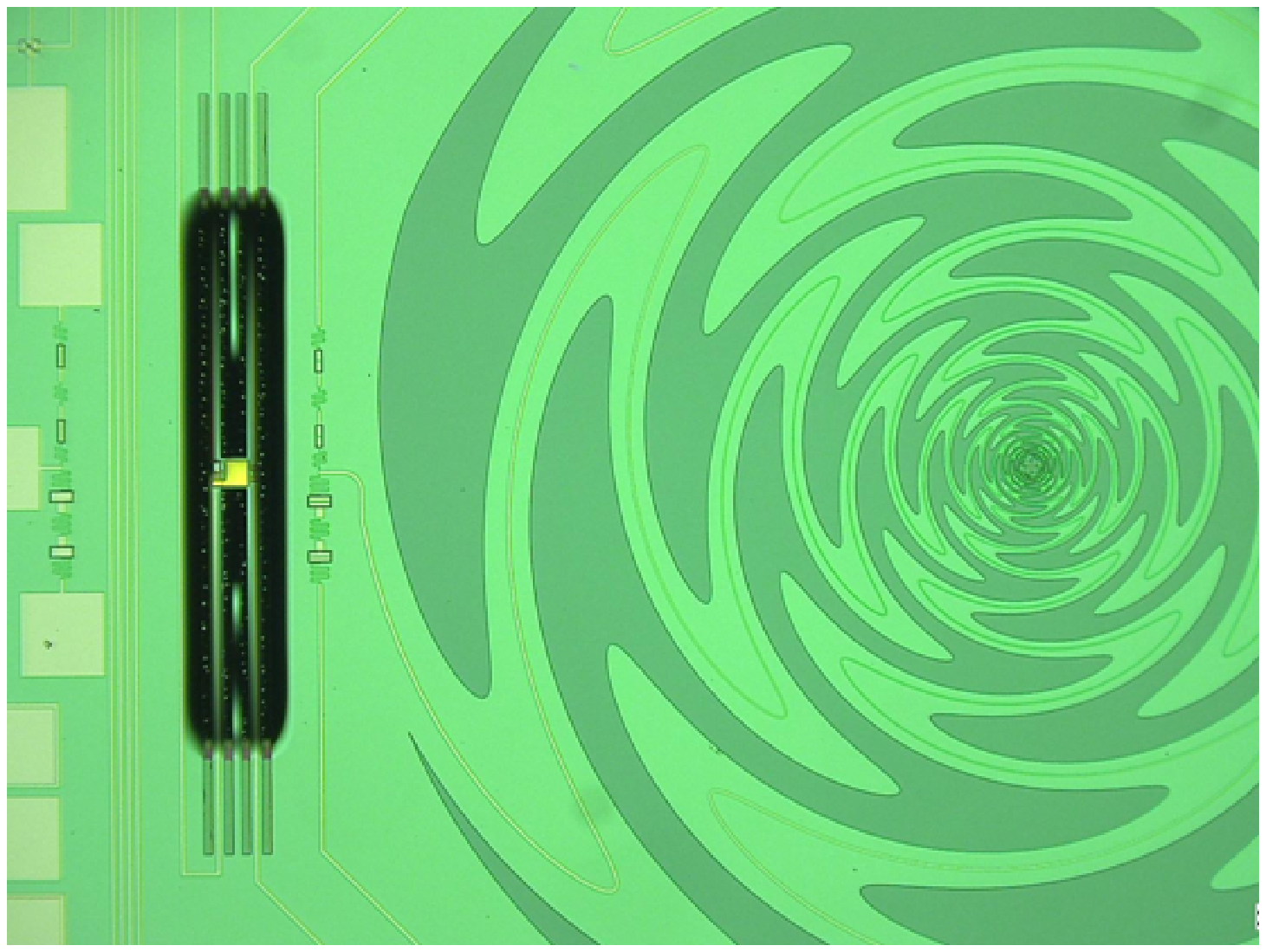}
\includegraphics[height=4cm,keepaspectratio]{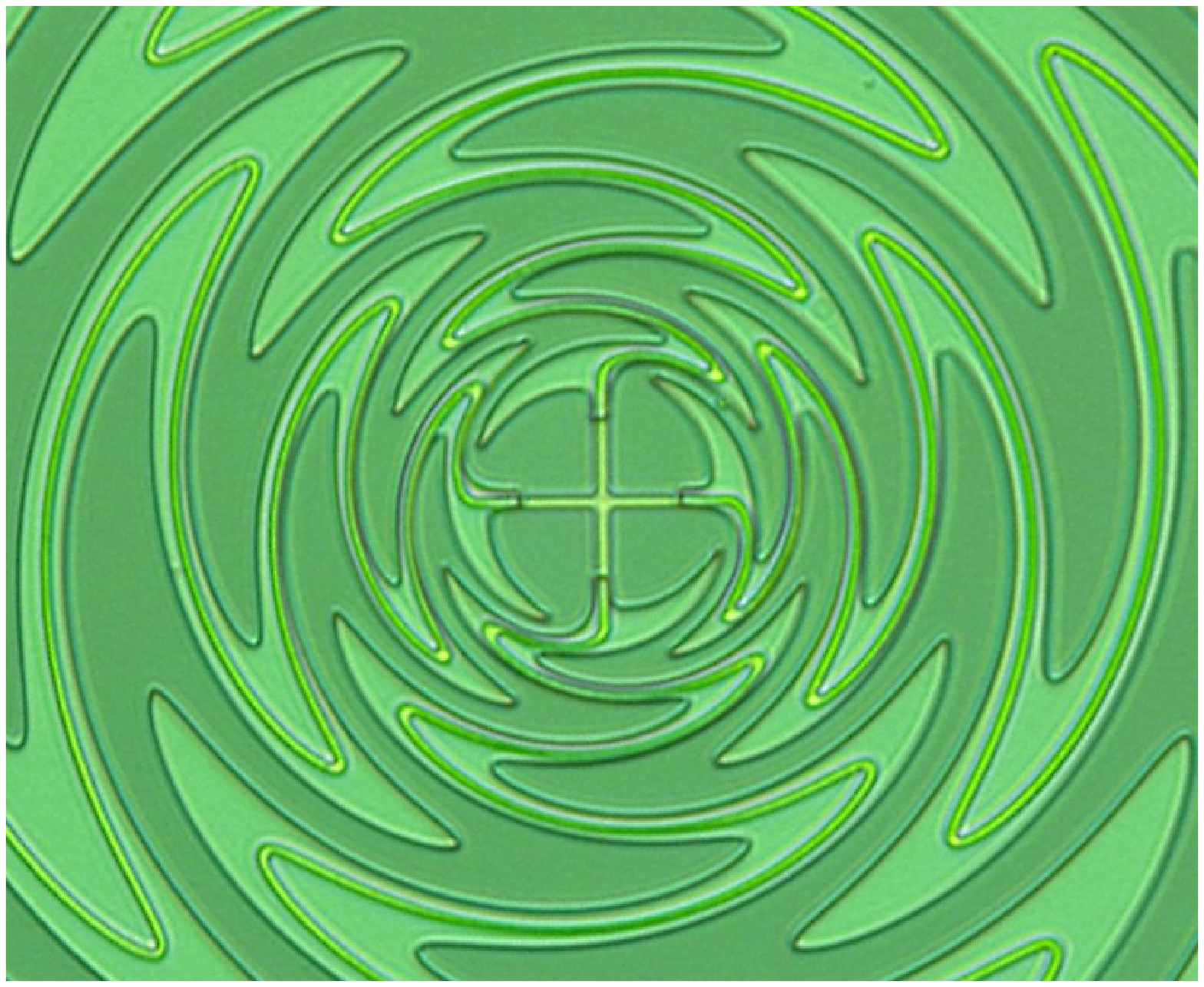}
\end{tabular}
\end{center}
\caption[example] 
{\label{fig:ant} 
(Left) side view of the silicon hemisphere lens with spacer. Spacer height is $0.38\times$ hemisphere's radius to place the antenna at the elliptical point. Rays drawn are schematic drawing.  (Center) photograph of the lumped diplexer pixel. The antenna is 1500~$\mu\mathrm{m}$ radius sinuous antenna. The black rectangular structure is the released TES bolometer. A lumped diplexer is between the antenna and the TES bolometer. (Right) Close-up of the center of the sinuous antenna. Crossed line is the microstripline. Two lines from orthogonal polarization crosses on the same layer. 
}
\end{figure} 
Figure \ref{fig:ant} shows our prototype pixel. We used an AR coated silicon hemisphere with a silicon extension to form a synthesized elliptical lens to increase the gain of the antenna \cite{Edwards,Filipovic} (Figure \ref{fig:ant}, left). The AR coating was a thermoformed Ultem1000 plastic shell epoxied on the silicon lens. The choice of the extension length of the synthesized lens is a compromise between the pixel's directivity and gausianity of the beam. We chose the extension length for the elliptical point ($0.38\times \mathrm{radius}$) to maximize directivity. The silicon half-space also increases front-to-back ratio to approximately 10:1 due to its high dielectric constant, and it makes the antenna more efficient. 

\begin{figure}
\begin{center}
\includegraphics[height=4cm,keepaspectratio] {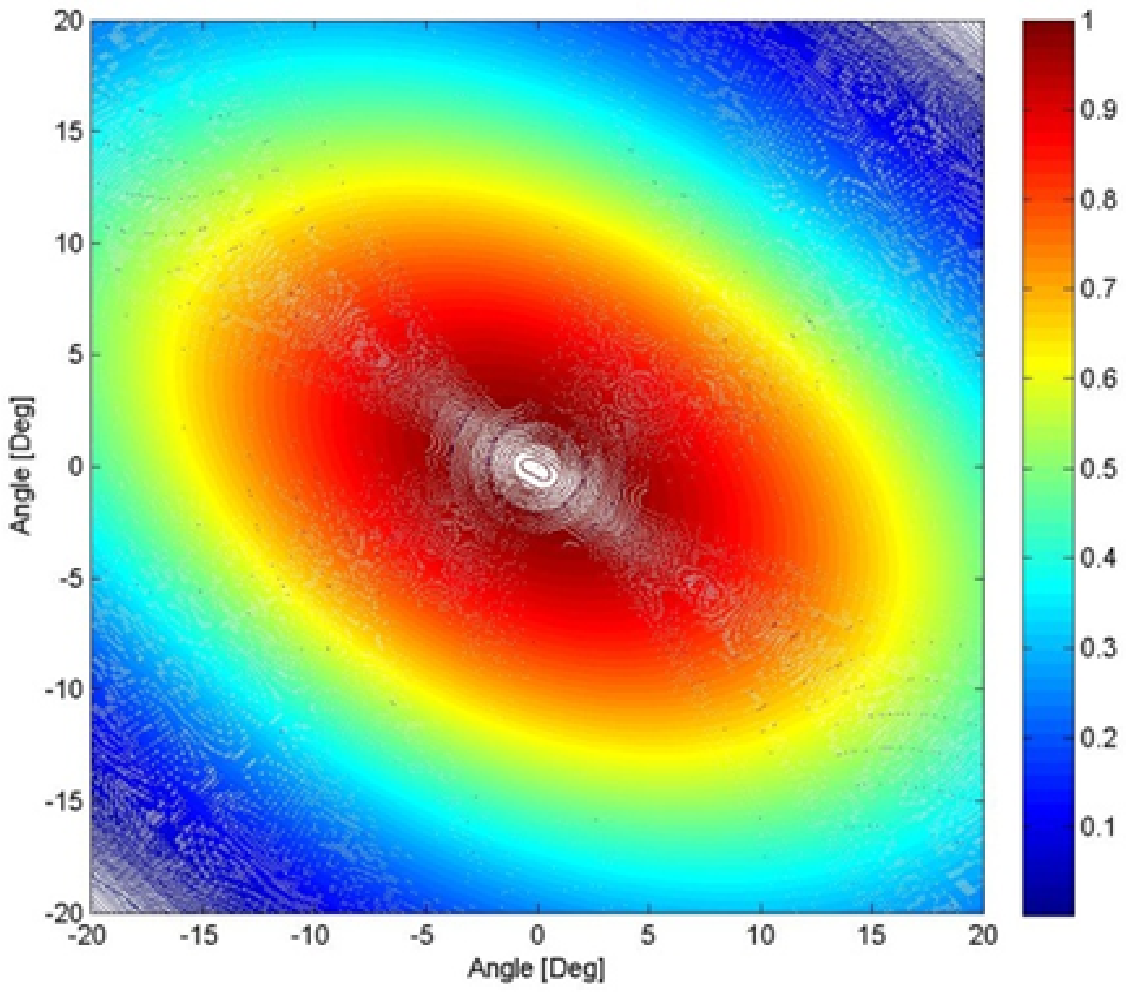}
\includegraphics[height=4cm,keepaspectratio] {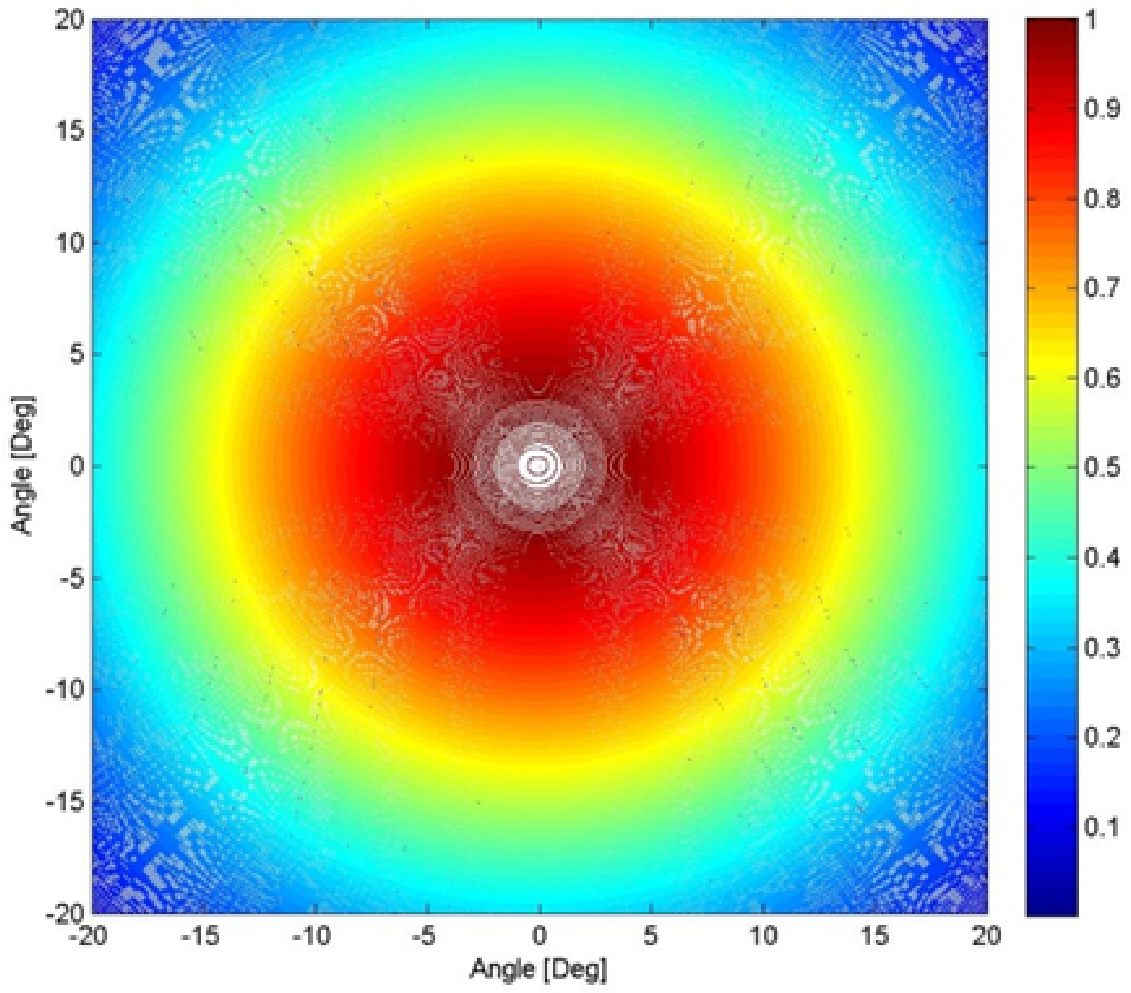}
\end{center}
\vspace{-3pt}
\caption{\label{fig:hfss}
Simulated beam at 80~GHz with a 540~$\mu\mathrm{m}$ radius sinuous antenna (left) and a 1500~$\mu\mathrm{m}$ radius sinuous antenna (right). The simulation was performed with the 3D EM software HFSS. Due to a long required computational time, the lens size was reduced to 6.35~mm diameter.
}
\end{figure}

For the antenna, we used a sinuous antenna. The sinuous antenna is a broadband antenna that has log-periodic and self-complementary structure and provides a stable input impedance over a  wide frequency range\cite{DuHamel}. Its self-complementarity is broken by the presence of the silicon half-space, but its impedance stays stable enough that antenna's reflection coefficient to an optimized constant real impedance transmission line stays below -10dB over its range\cite{Edwards}. The sinuous antenna is favored over other log-periodic antennas for its sensitivity to linear polarizations. It also has a small polarization axis rotation as a function of frequency\cite{Edwards}.  We want to keep the logarithmic expansion coefficient, frequency of repetitive structure, ($\tau$) as close to unity as possible to keep the impedance fluctuation and polarization axis rotation as small as possible. A smaller expansion coefficient requires a smaller antenna and microstrip features. We chose $\tau = 1.3$ given micro-fabrication limitations. We chose the inner and outer radii of the antenna to be 15~$\mu\mathrm{m}$ and 1500~$\mu\mathrm{m}$ respectively. With an expansion factor of 1.3, the 1500~$\mu\mathrm{m}$ radius fits 16 repetitions ({\it cells}) of the antenna. A four-arm, self-complementarity sinuous antenna on silicon radiates at 360~$\mu\mathrm{m}$ radius for 70~GHz, the lowest frequency we use. From the results of 3D EM simulation with HFSS, we found out that a larger outer radius was required to obtain stable input impedance and round beam shape at our operating frequencies (Figure \ref{fig:hfss}). This result was also observed experimentally by testing samples with antennas of different sizes. We will show results from a distributed diplexer, a distributed triplexer and a channelizer, which had a sinuous antenna with a radius of 540~$\mu\mathrm{m}$. These result demonstrate the multiplexing technique. We will then show the beam performance of an antenna with a 1500~$\mu\mathrm{m}$ radius using a lumped diplexer.

\begin{figure}

\begin{center}
\includegraphics[height=4cm,keepaspectratio] {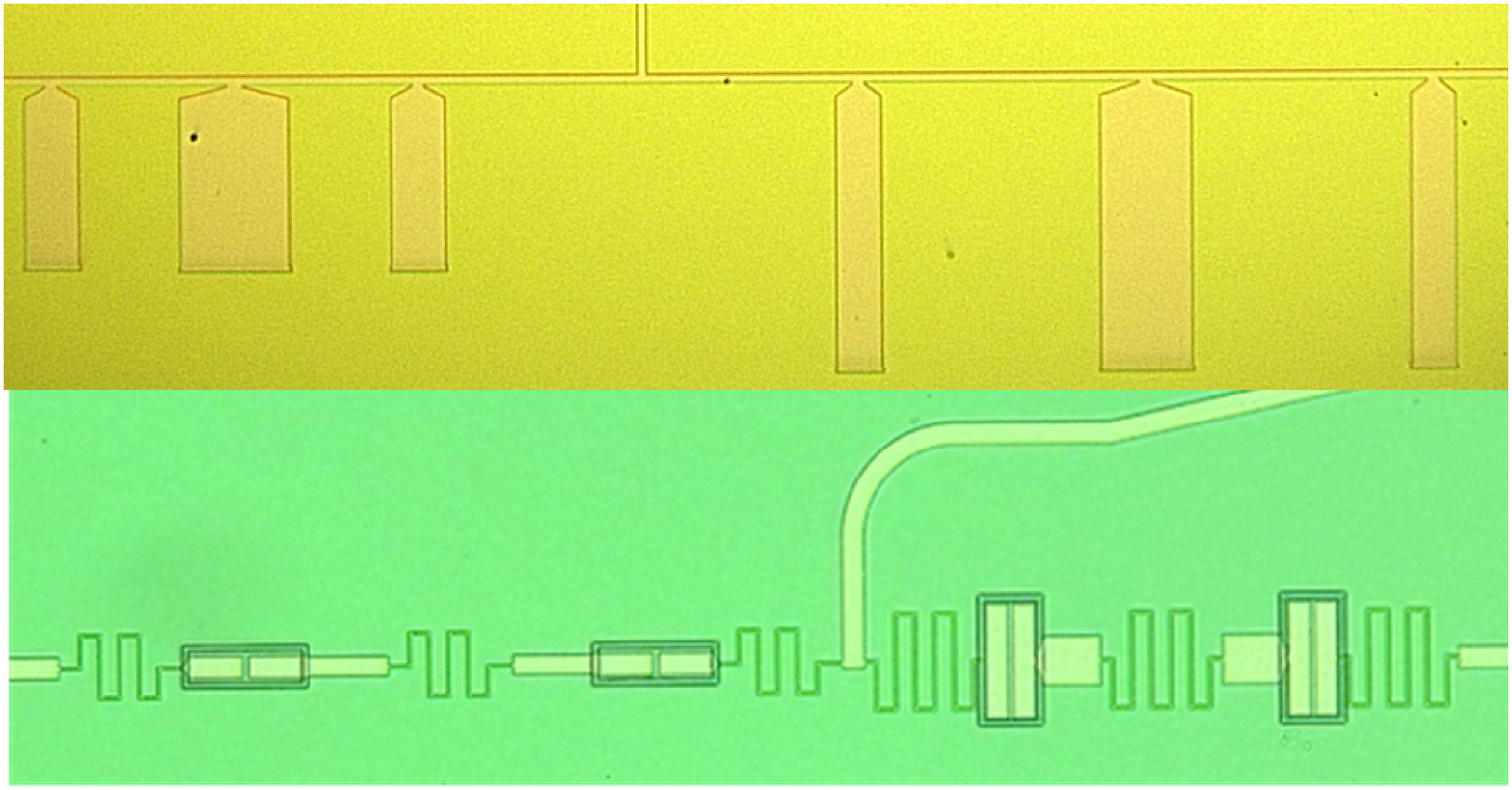}
\includegraphics[height=4cm,keepaspectratio] {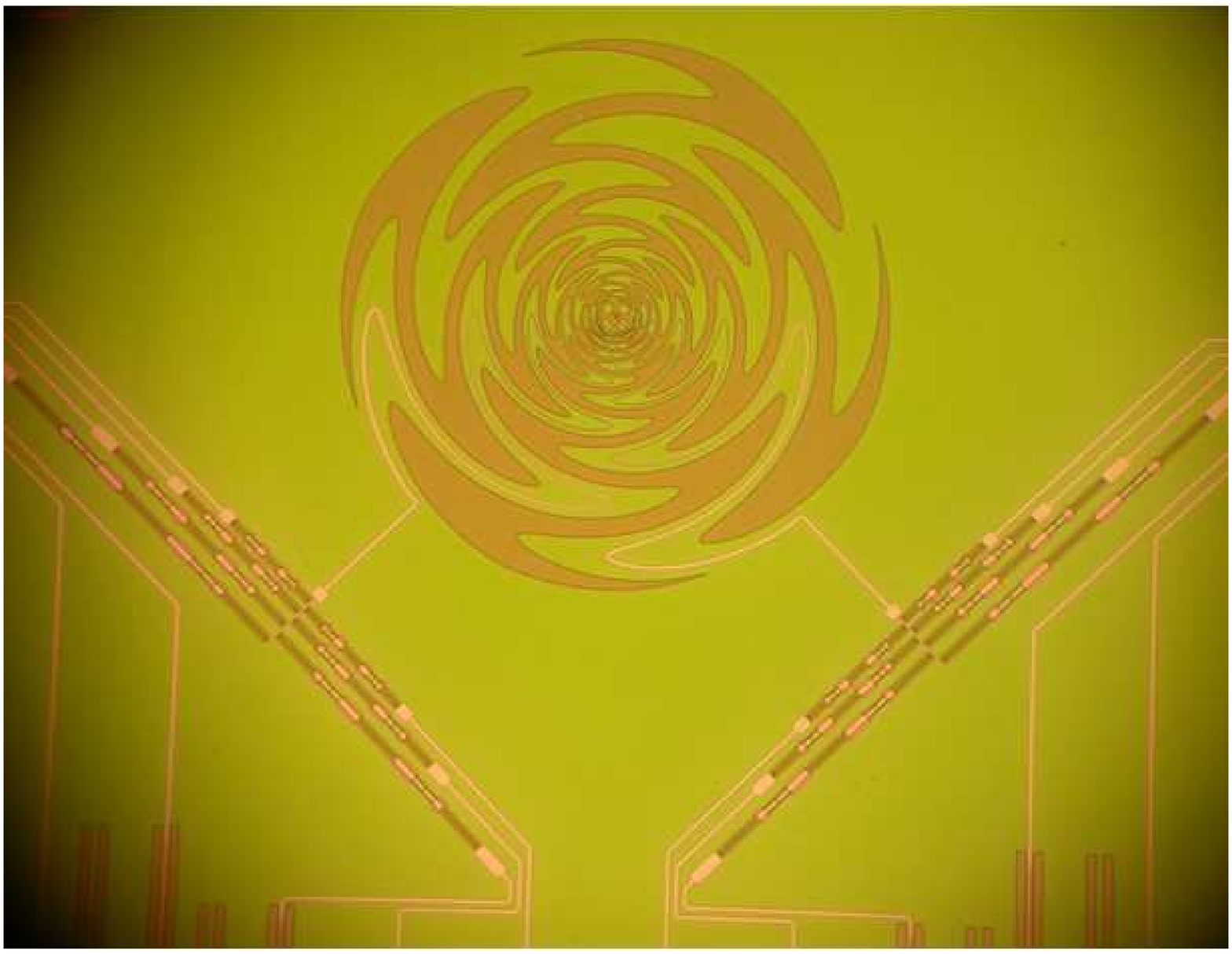}
\end{center}
\vspace{-3pt}
\caption{\label{fig:filter} 
(Left top) photograph of a distributed diplexer. Stubs are shorted to the ground plane at the ends. A distributed triplexer has similar appearance except there is one more filter branch attached to the junction. (Left bottom) photograph of the lumped diplexer. The meander structure is the inductor created from a narrow (2~$\mu\mathrm{m}$) microstrip line. The rectangular structures are capacitors. (Right) photograph of the channelizer filter attached to a 540~$\mu\mathrm{m}$ radius sinuous antenna. The spiral structure is the sinuous antenna, and each leg of the V-shaped structure is the channelizer filter for one of the linear polarizations. 
}
\end{figure}

Power from the antenna is coupled to microstrip lines to allow RF frequency-selection with on-chip filters prior to detection at the bolometers. The metalized {\it arms} of the sinuous antenna are used as the ground plane of the microstrip line so that we can bring microstrip line to the center of the antenna without interfering with the antenna. This scheme allows a planar design, and this allows large array fabrication with lithography technique. To simplify the fabrication process, microstrip lines for orthogonal polarizations cross at the center on the same layer (Figure \ref{fig:ant}, right). They do not couple to each other since odd-mode excitation of the antenna creates a virtual short at the intersection.

For the multiplexing method, we explored four schemes of paritioning: distributed diplexers, distributed triplexers, lumped element diplexers and channelizers with seven frequency bands (Figure \ref{fig:filter}). Both the diplexer and triplexer were developed with terrestrial observations in mind. We strategically placed bands between atmospheric emission lines to avoid detecting photons from the hot atmosphere. Both the distributed element filter and lumped element filter designs were based on a 0.5~dB ripple three-pole Chebyshev band pass filter, as a compromise between loss due to the filter elements and roll-off speed\cite{Arnold}. The quarter-wavelength stub filter consists of three stubs of shorted $\lambda / 4$ lines separated by $\lambda / 4$ lengths of microstrip lines. For a lumped filter design, we designed an inductor with short lengths of narrowed (2~$\mu\mathrm{m}$) microstrip line. For a capacitor, we designed a parallel plate capacitor with two layers of niobium separated by silicon dioxide. For calculating the required values of inductor and capacitor, we followed O'Brient {\it et al.} and Kumar {\it at al/} \cite{OBrient, Kumar}. Filters for each frequency were independently optimized usint the Sonnet EM 2.5 dimension simulation package. The effect of superconducting kinetic inductance was account for using a surface impedance of 0.11pH/sq\cite{Kerr}. The distances between the microstrip junction and each filter were adjusted until -20~dB isolation between frequency bands was obtained.

The channelizer was designed with a satellite mission in mind, where atmospheric lines are not a problem. Therefore, optimal performance is obtained by partitioning bands that are adjacent to each other. In the channelizer, lumped band pass filters are spaced in a log-periodic frequency schedule, and attached to the transmission line from high to low frequency. Within the filter's resonant band pass,the  signal is transmitted through the filter to its associated bolometer, while other frequencies continue on the main transmission line. Galbraith and Rebeiz realized a three-pole channelizer at lower frequencies using surface mount lumped components\cite{Galbraith}. We adapted this design to thin film fabrication\cite{OBrient}. We made a seven channel channelizer to cover the 70 to 220 GHz range.

The signals are detected by superconducting Transition Edge Sensor (TES) bolometers. Microstrip lines from the antenna are terminated by a 20~$\Omega$ resistor made of aluminum-titanium bilayer. We made a dual-transition TES with 1~$\mathrm{\Omega}$ aluminum-titanium bilayer and 10~$\mathrm{\Omega}$ aluminum in series. We used the same metal layer for the termination resistor and the TES thermistor to simplify fabrication. The  superconducting transition temperature of the aluminum-titanium bilayer was tuned to be around 0.5~Kelvin by changing the thickness of the titanium. Aluminum's superconducting transition is assumed to be 1.2~Kelvin. This difference in Tc allows us to operate the TES bolometer over two orders of magnitude difference in receiving power, such that we can use aluminum-titanium transition for CMB observations while the aluminum transition is used in lab tests. Between the termination resistor and TES, we left layers of aluminum-titanium as well as a 1.5 thick micron gold. This provides additional heat capacity, allowing detector time constants to be tuned to match the readout bandwidth. The bolometer structure is supported on four 0.8 micron thick and about five hundred micron long nitride legs to provide thermal isolation from the rest of the chip. A summary of the layers of the device and the fabrication processes is shown in Table \ref{fab}.
\begin{table}
\begin{center}
  \begin{tabular}{ l r r r }
	\hline
	Material & Thickness [$\mu\mathrm{m}$] & Deposition & Etch \\ \hline
	Silicon & 500 & test-grade 6in. & $\mathrm{XeF}_2$ \\
	Silicon Dioxide & 0.05 & thermal growth  & $\mathrm{CF}_4$ RIE \\
	Low Stress Nitride & 0.7 & LPCVD & $\mathrm{CF}_4$ RIE\\
	Niobium & 0.3 & DC Mag. Sputter & $\mathrm{CF}_4$/$\mathrm{O}_2$ RIE \\
	Silicon Dioxide & 0.5 & $350^{\circ}$C PECVD & $\mathrm{CHF}_3$/$\mathrm{O}_2$ RIE \\
	Niobium & 0.3 & DC Mag. Sputter & $\mathrm{CF}_4$/$\mathrm{O}_2$ RIE \\
	Silicon Dioxide & 0.2 & $350^{\circ}$C PECVD & $\mathrm{CHF}_3$/$\mathrm{O}_2$ RIE \\
	Niobium & 0.3 & DC Mag. Sputter & $\mathrm{CF}_4$/$\mathrm{O}_2$ RIE \\
	Al/Ti Bilayer &  0.04/0.08 & DC Mag. Sputter & Pre-mixed Wet Etch/$\mathrm{SF}_6$ RIE \\
	Gold & 1.5 & Evaporation. & Photo.Res. Lift-off \\
	\hline
  \end{tabular}
\end{center}
\vspace{-3pt}
\caption{Summary fabricated device layers and fabrication processes. The layers are in the order of preparation.}
\label{fab}
\end{table}
\section{TEST SETUP}\label{sec:test}
\begin{figure}
\begin{center}
\begin{tabular}{c}
\includegraphics[height=6cm,keepaspectratio]{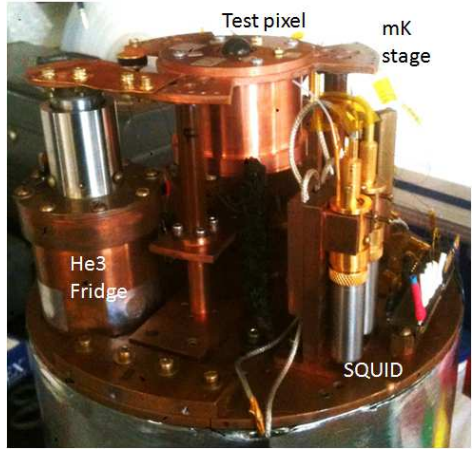}
\end{tabular}
\end{center}
\caption[example] 
{\label{fig:setup} 
Photograph of the milli-kelvin stage of the IR Labs dewar. Homemade 3He fridge sit on the left side of the dewar. Milli-Kelvin stage is isolated from 4~Kelvin plate by thin vespel leg approximately 2~inch long and 0.010~inch thick. Silicon lens of the test pixel is visible at the center of milli-kelvin stage. Quantum Design DC SQUID readout was used to read out TES bolometer, that is visible on the right. 
}
\end{figure} 

We tested the prototype pixel in an 8~inch IR Labs dewar. We modified the dewar by adding a 4~inch diameter optical window made from Zotefoam. For infrared filters, two layers of 0.125~inch thick expanded teflon and a metal mesh low pass filter with $18\mathrm{cm}^{-1}$ cut off are anchored to a liquid nitrogen temperature. Two metal mesh low pass filters with cut off at $14\mathrm{cm}^{-1}$ and $12\mathrm{cm}^{-1}$ are mounted at liquid helium buffer to further reduce the optical loading. The pixel stage is isolated from the liquid helium buffer with thin walled vespel tubes. The stage is cooled to 0.25~Kelvin with homemade 3He adsorption fridge. We mounted the test pixel behind the 14mm diameter hemispherical silicon lens with 2mm thick flat silicon spacer. The test pixel was fabricated on top of 0.675~mm thick silicon, thus combination of spacer and the test pixel locate the antenna right at the elliptical focus.
The TES bolometer is DC voltage biased with $0.02\Omega$ of shunt resistor in parallel with the bolometer. Current through the bolometer is read out by commercially available laboratory DC SQUID from Quantum Design with its input inductor coil in series with the bolometer. 

We produced beam maps of the pixel by scanning 0.25~inch diameter temperature modulated source (liquid nitrogen soaked eccosorb and room temperature eccosorb) 10~inches away from the antenna. We scanned 3~inch $\times$ 3~inch patch with step size of 0.125~inch on motorized XY stage. We measured the response of the pixel to a linear polarized source by rotating wire grid polarizer between the pixel and the temperature modulated source. We measured spectra of the device using Michaelson Fourier transform spectrometer (FTS). The FTS uses temperature modulated source with 1200~Kelvin ceramic heater and 300~K eccosorb. Mirrors are 6~inch by 6~inch large in cross-section. Beam splitter was 0.010~inch thick mylar that has a beam splitter minima at 360~GHz. We focused the output of the FTS onto the pixel using ultra high molecular weight polyethylene lens. The efficiency of the device was measured with beam filling temperature modulated source. For a single moded antenna detector, the power difference between two temperature source is $k_B\Delta T \Delta \nu$ in the Rayleigh-Jean limit. Here $k_B$ is the boltzmann constant, $\Delta T$ is the difference in temperature of modulated source. We used liquid nitrogen soaked eccosorb and room temperature eccosorb for $\Delta T = 223$~Kelvin. $\Delta \nu$ is the integrated bandwidth of the peak normalized spectrum measured with FTS. We divide power received on detector with $k_B\Delta T \Delta \nu$ to measure an end-to-end efficiency which includes dewar loss.

\section{RESULTS}\label{sec:result}
\begin{figure}
\begin{center}
\begin{tabular}{c}
\includegraphics[height=5cm,keepaspectratio]{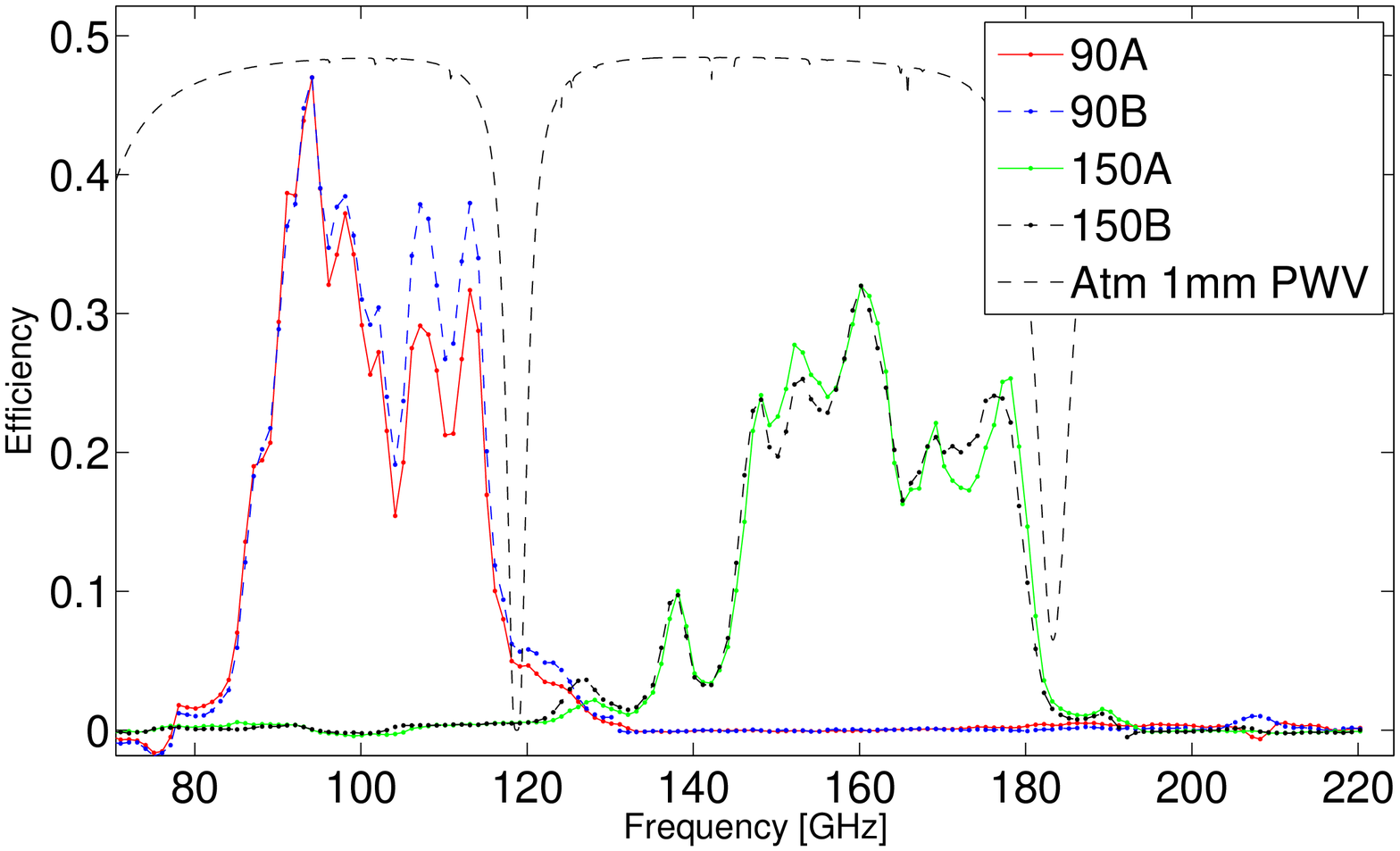}
\includegraphics[height=5cm,keepaspectratio]{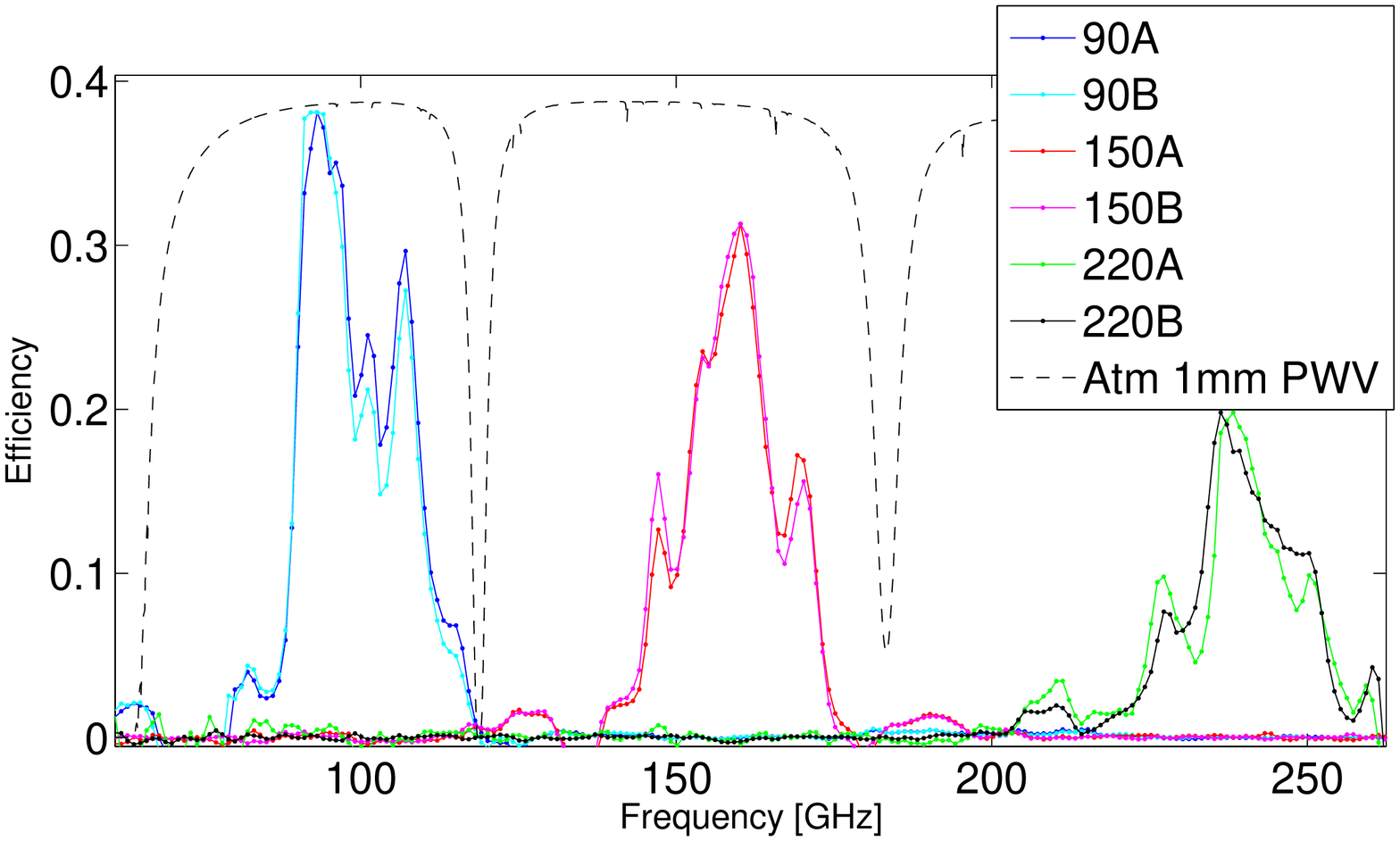}
\end{tabular}
\end{center}
\caption[example] 
{\label{fig:specdist} 
Spectrum of a distributed diplexer (left) and a distributed triplexer (right). {\it A} and {\it B} refers to two orthogonal linear polarization channels. Peaks are normalized to the measured optical efficiency. See Table \ref{result} for details. Dotted black lines are expected transmission through 1mm PWV of atmosphere.}
\end{figure} 

\begin{figure}
\begin{center}
\begin{tabular}{c}
\includegraphics[height=5cm,keepaspectratio]{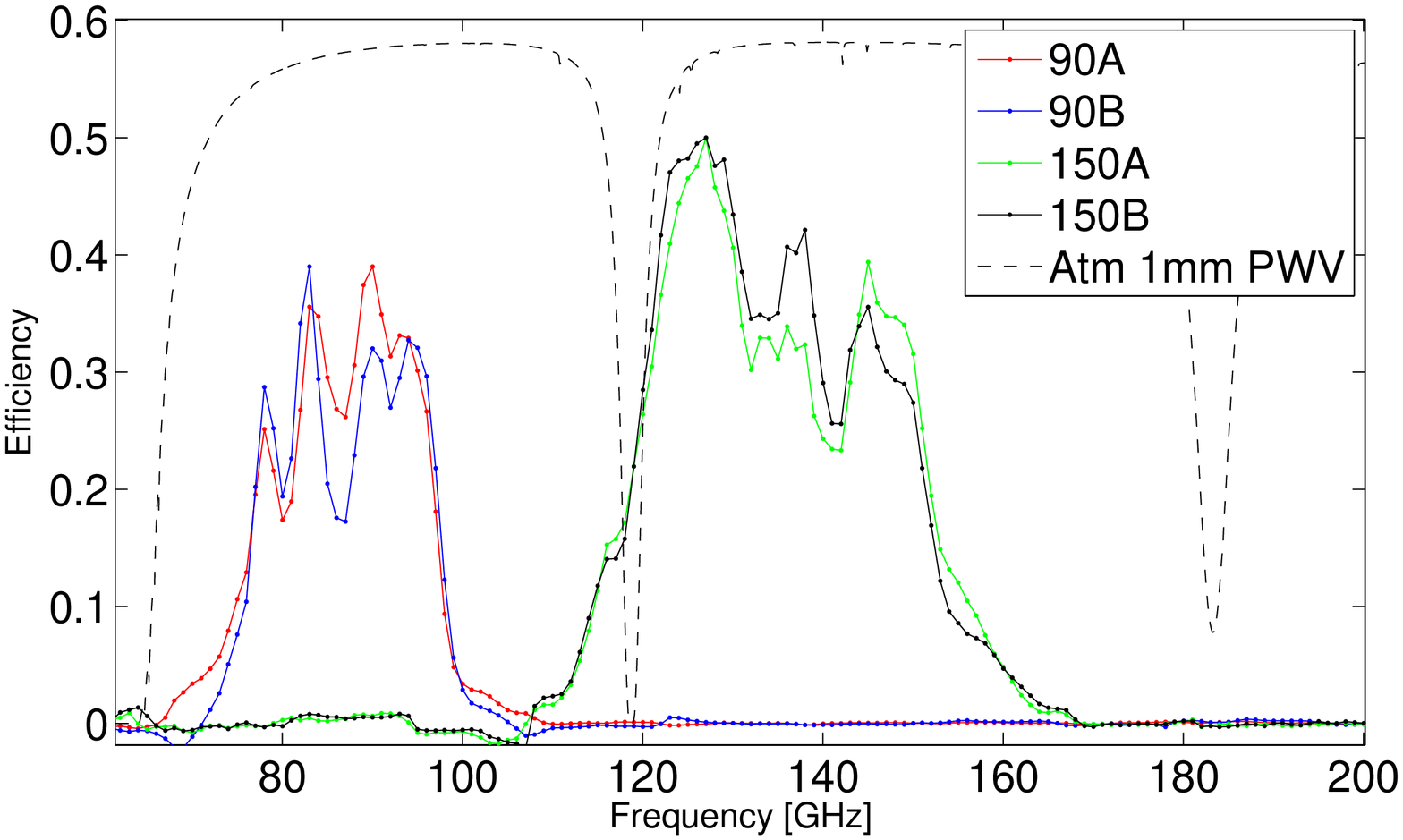}
\includegraphics[height=5cm,keepaspectratio]{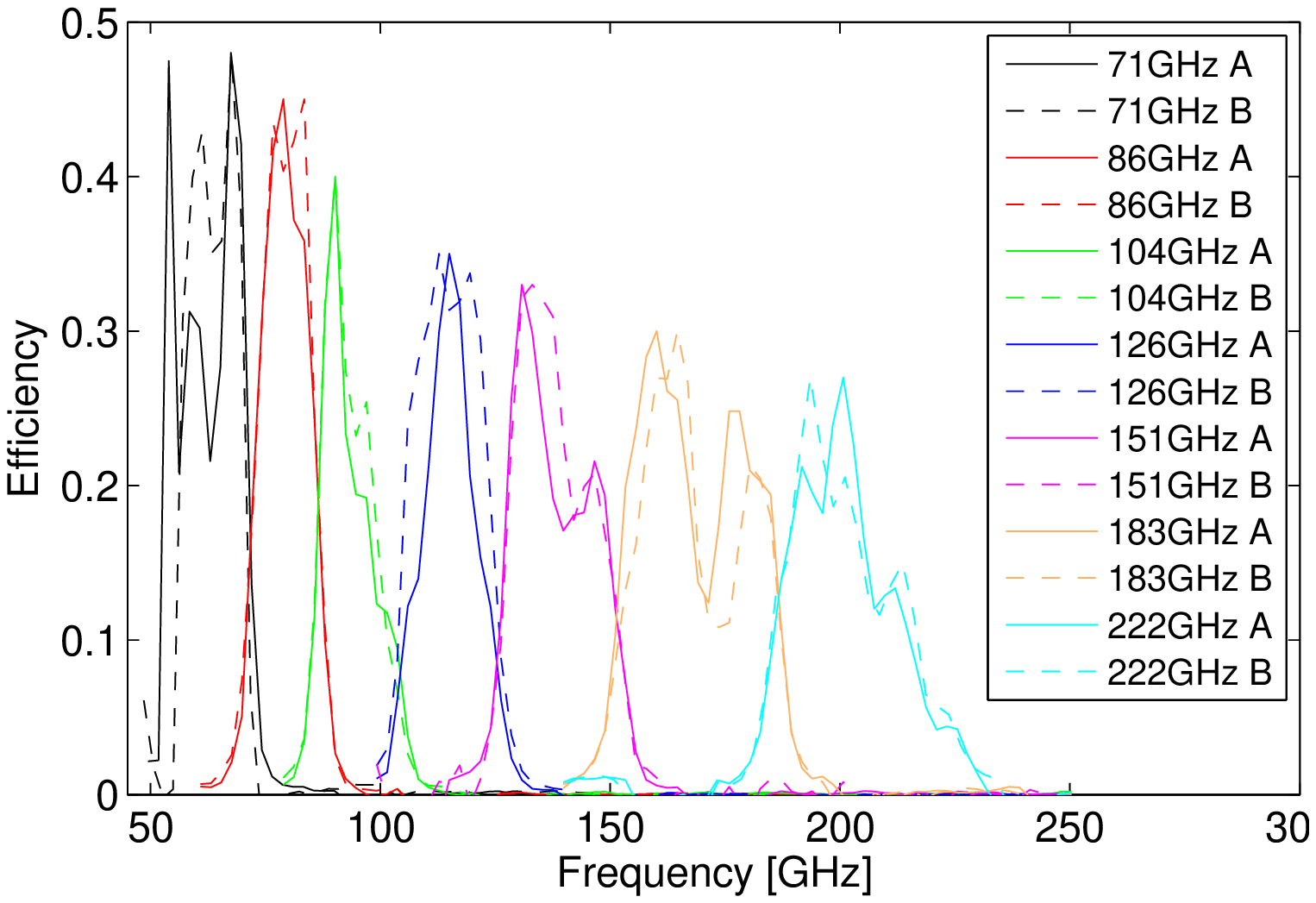}
\end{tabular}
\end{center}
\caption[example] 
{\label{fig:speclump} 
Spectrum of a lumped diplexer (left) and a channelizer (right). {\it A} and {\it B} refers to two orthogonal linear polarization channels. Peaks are normalized to measured optical efficiency. See Table \ref{result} for details. Dotted black lines are expected transmission through 1mm PWV of atmosphere.}
\end{figure} 

\begin{figure}
\begin{center}
\begin{tabular}{c}
\includegraphics[height=5cm,keepaspectratio]{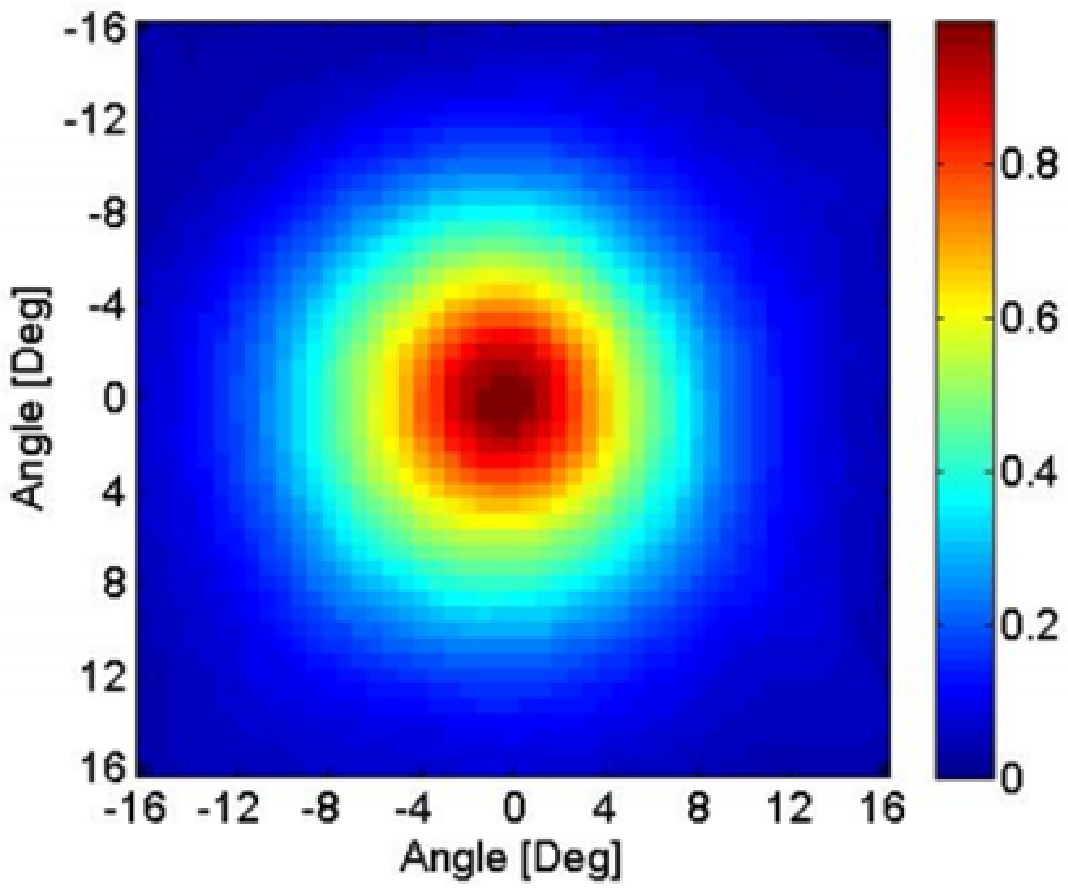}
\includegraphics[height=5cm,keepaspectratio]{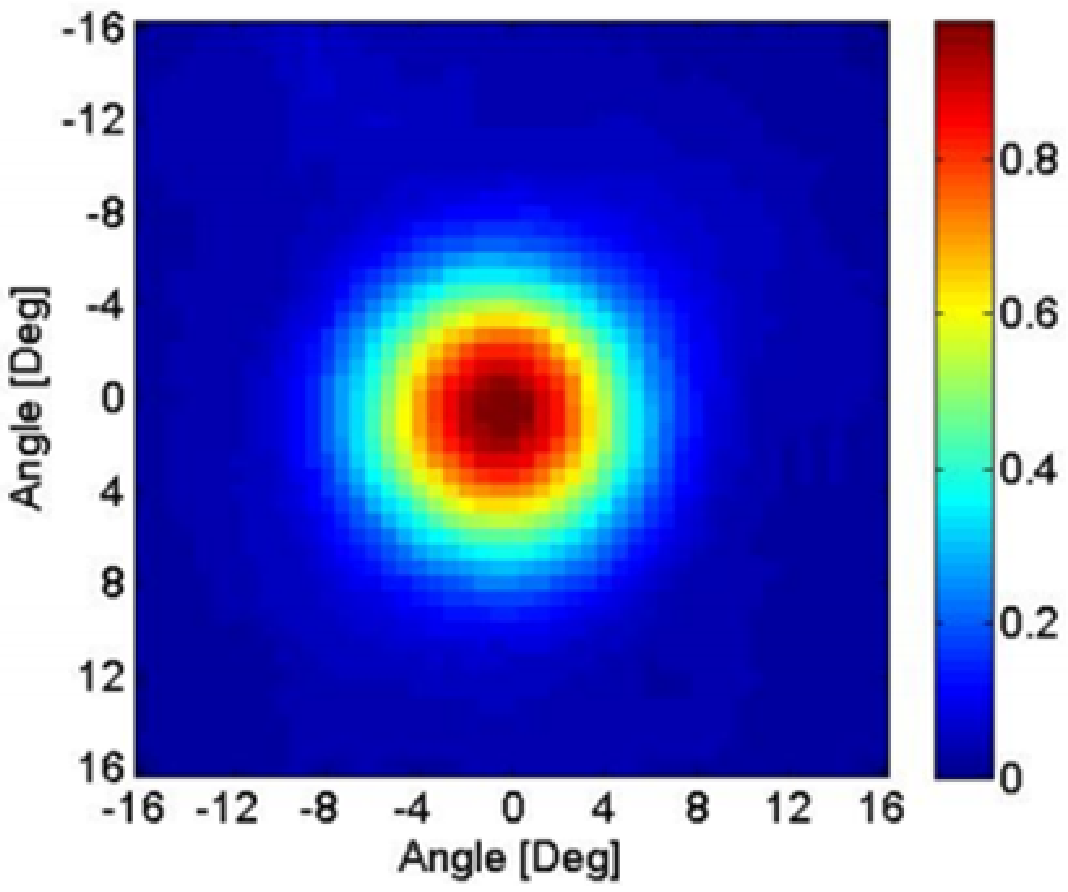}
\end{tabular}
\end{center}
\caption[example] 
{\label{fig:beammap} 
Beam map measurement from a lumped diplexer. Plots were peak normalized. Lumped diplexer was the only test structure that was attached to large (1500~$\mu\mathrm{m}$ radius) sinuous antenna. Ellipticity was calculated by fitting an ellipse on the -3~dB contour and used the definition $\epsilon = (a-b)/(a+b)$, where $a$ and $b$ are the major and the minor axes. Ellipticities were $1.1\%$ and $1.3\%$ for 90GHz and 150GHz  band respectively. }
\end{figure} 

\begin{figure}
\begin{center}
\begin{tabular}{c}
\includegraphics[height=5cm,keepaspectratio]{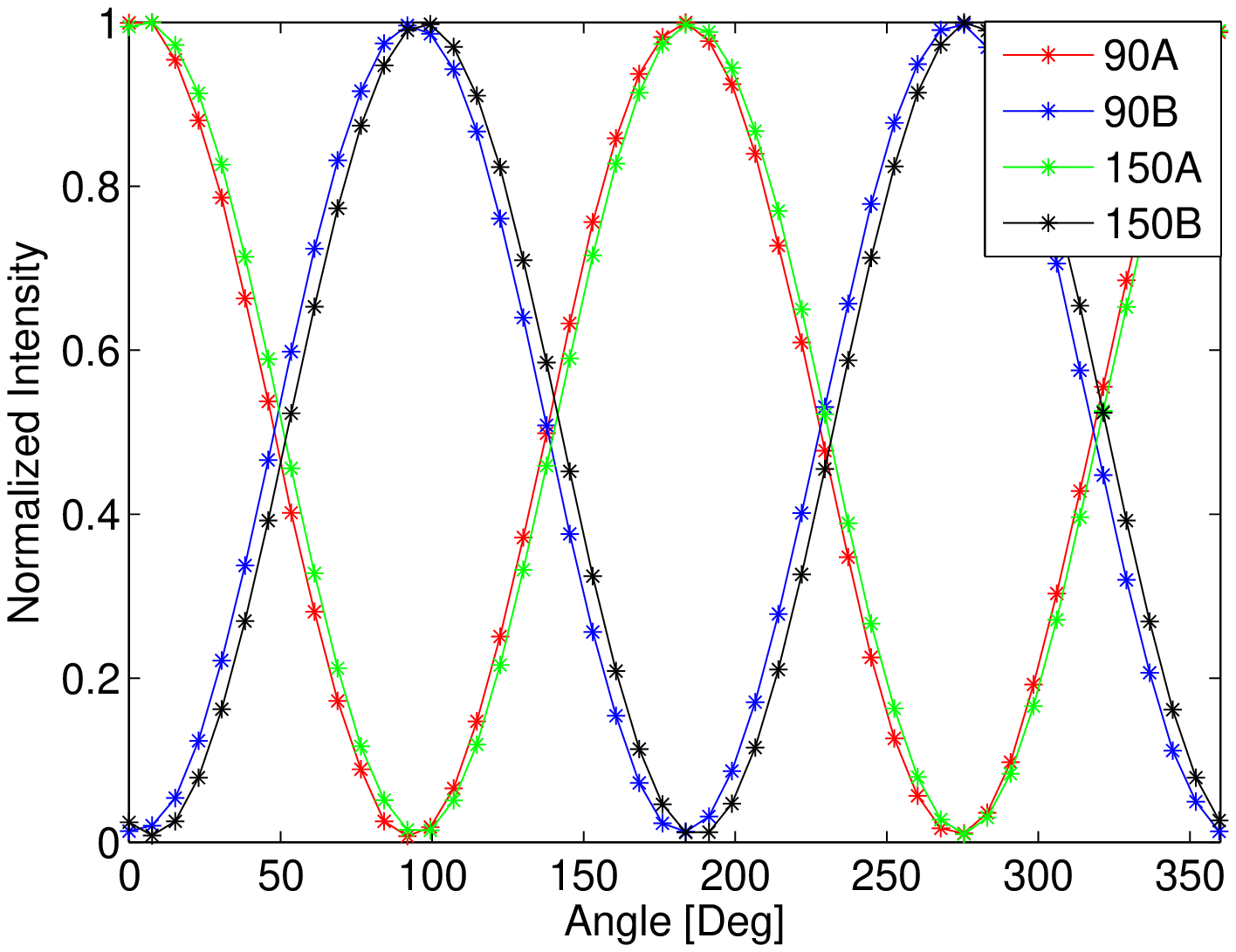}
\includegraphics[height=5cm,keepaspectratio]{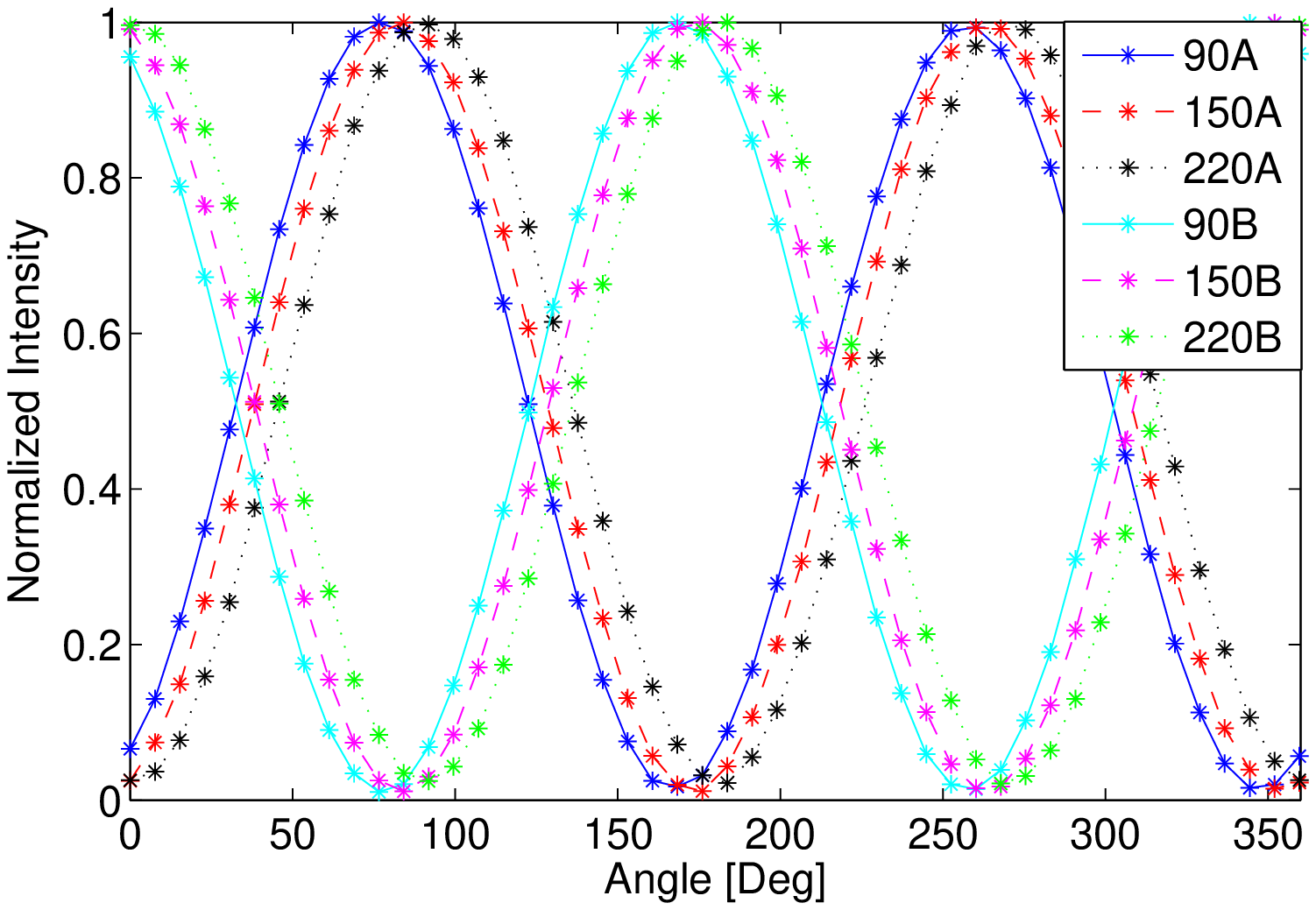}
\end{tabular}
\end{center}
\caption[example] 
{\label{fig:polarization} 
Responses of the lumped diplexer (left) and distributed triplexer (right) to a linearly polarized source as a function of relative angle between antenna and the polarizer. Plots were peak normalized prior to fitting by sum of a sine function and a constant. Cross-pol for each channels are summarized in Table \ref{result}.}
\end{figure} 

\begin{table}
\begin{center}
  \begin{tabular}{ l r r r r r r }
	\hline
	Filter Type & Ant rad [$\mu\mathrm{m}$]& $\nu_0$ [GHz] & $\Delta \nu$ [GHz] & Opt Eff& Ellipticity & Cross-pol \\ \hline
	Lumped Diplexer Low & 1500 & 87 & 17.0 & 39\% & 1.1\% & $<0.3$\%\\
	Lumped Diplexer Mid & 1500 & 135 & 26.4 & 50\% & 1.3\% & $<1.6$\%\\
	Stub Diplexer Low & 540 & 101 & 20.2 & 47\% & 4.9\% & $<2.3$\%\\
	Stub Diplexer Mid & 540 & 162 & 26.2 & 32\% & 1.0\% & $<1.6$\%\\
	Stub Triplexer Low & 540 & 100 & 16.6 & 38\% & 3.0\% & $<2.5$\%\\
	Stub Trilexer Mid & 540 & 158 & 17.7 & 31\% & 1.5\% & $<2.1$\%\\
	Stub Trilexer High & 540 & 239 & 19.6 & 20\% & 4.0\% & $<4.3$\%\\
	\hline
  \end{tabular}
\end{center}
\vspace{-3pt}
\caption{Summary from one of the polarizations of each diplexer and triplexer. $\nu_0$ is the center frequency of the band and $\Delta \nu$ is integrated bandwidth. Cross-pol values are upper limit value as we expect leakage from wire-grid}
\label{result}
\end{table}

The results from lab measurements are summarized in Table \ref{result}. To analyze the spectra, the interferogram from the FTS was apodized with triangular window function prior to the Fourier transformation. Then the spectrum was divided by analytical beam splitter function to remove its effect. The resulting spectra from distributed diplexer and distributed triplexer are shown in Figure \ref{fig:specdist}, and the spectra from lumped filter diplexer and channelizer are shown in Figure \ref{fig:speclump}. Peaks of the spectra were normalized to a measured optical efficiency of each band. The results show that we successfully  partitioned a broadband signal into 2, 3 and 7 bands with matching band shape for orthogonal polarizations. We did not know the material properties ahead of the test pixel fabrication to get the band location correct. The \Pb\ CMB experiment successfully tuned their band location of the filters by making correction to the filter design with feedback from lab measurements \cite{Arnold}. Our filter can be tuned in the same way in future fabrications. 

The beam maps from a lumped diplexer pixel with a 1500~$\mu\mathrm{m}$ radius sinuous antenna are shown in Figure \ref{fig:beammap}. We characterize the beam property by fitting ellipse at -3~dB contour and define ellipticity as $\epsilon = (a-b)/(a+b)$, where $a$ and $b$ are the major and the minor axes of the ellipse, respectively. Ellipticity from large antenna is $1.1\%$ and $1.3\%$ for 90~GHz and 150~GHz  band respectively. At this level, we were limited by measurement systematics. Comparing 1500~$\mu\mathrm{m}$ radius antenna measurements with the 540~$\mu\mathrm{m}$ radius antenna, we clearly see the improvement in beam ellipticity for the large antenna at low frequencies, as predicted from 3D EM simulation. 

Polarization measurement results are shown in Figure \ref{fig:polarization} and tabulated at Table \ref{result}. We expect the wiregrid to have approximately $1\%$ leakage, thus we are limited by systematics for low cross-pol measurement. We also see the that polarization performance for the lower frequency improves with the larger antenna.

\section{ANTI-REFLECTION COATING}\label{sec:ar}
\begin{figure}
\begin{center}
\begin{tabular}{c}
\includegraphics[height=6cm,keepaspectratio]{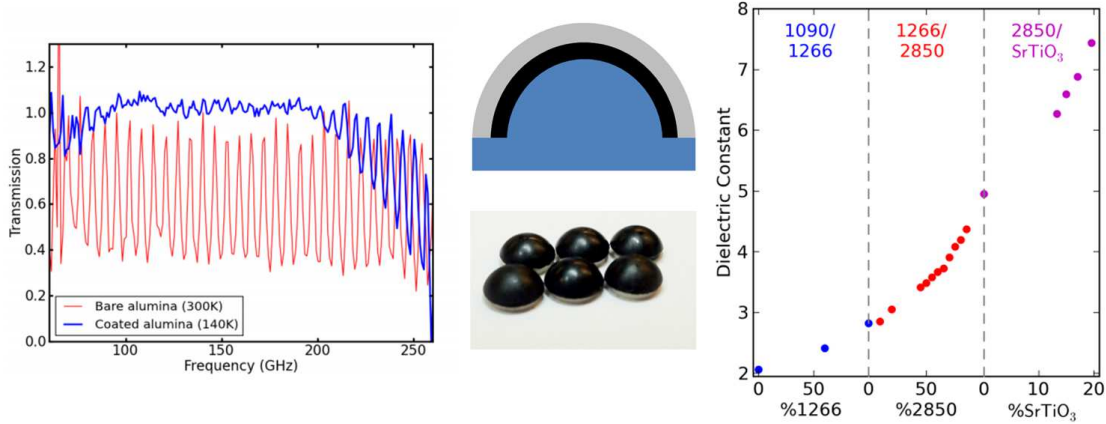}
\end{tabular}
\end{center}
\caption[example] 
{\label{fig:ar} 
(Left) Transmission of 2-layer AR coated alumina (blue) and uncoated alumina sample (red) measured with an FTS. Fabry-Perot fringes disappears over $70\%$ of bandwidth for the AR coated sample. (Center top) Cut-away drawing showing how two layer coatings are applied. For a silicon lens, the inner layer (black) would be Stycast 2850FT ($\epsilon = 4.95$) and outer layer would be Stycast 1090 ($\epsilon = 2.05$). (Center bottom) alumina lens coated with 2-layers of AR coating. Inner layer is Stycast 2850FT and the outer layer is Stycast 1090. (Right) Plot of measured dielectric constant of the material and various epoxy and filler mix. Left region is a mix of S 1090 and Stycast 1266, center region is a mix of Stycast 1266 and Stycast 2850FT and right section is the mix of Stycast 2850FT and $\mathrm{SrTiO_3}$ powder. Percentage on the x-axis is the percentage of mass of second material in the mix. 
}
\end{figure} 
One of the optical challenges  of building a multi-chroic system is the production of broadband anti-reflection (AR) coatings. Since our pixel uses silicon lens that has a high dielectric constant ($\epsilon = 11.7$) there would be a $30\%$ reflection without an AR coating. Also in the future CMB experiments such as \Pb-2, reimaging lenses will be fabricated from alumina which also has a high dielectric constant $\epsilon = 9.6$. Solving the broadband AR coating issue is one of the keys to make the multi-chroic technology viable. 

The bandwidth of the coating increases with the number of correctly tuned layers, but absorptive loss will also increase. Thus we want to keep the number of layers as low as possible and still provide enough bandwidth. The coating must stay on the lens at cryogenic temperatures. For millimeter wave application, the thickness of the AR coating will be order 100~$\mu\mathrm{m}$. Thus it is too thin to easily machine and too thick for vapor deposition. We would also like to be able to tune the dielectric constant of the coating to obtain optimal coating performance.

We calculated that a 2-layer coating with dielectric constants of 2 and 5 with thicknesses of one-quarter wavelength at the of center frequency will give sufficient bandwidth to cover 90~GHz and 150~GHz bands. To cover 90,150 and 220~GHz bands simultaneously, we need a 3-layer coating with dielectric constants of 2,4, and 7 with thickness of a quarter of wavelength at the center frequency. We chose to use mixtures of various types of epoxies (Stycast 1090, Stycast 1266, Stycast 2850FT) and filler ($\mathrm{SrTiO_3}$) which can be molded into appropriate AR coating layers. 

First we measured the dielectric constant of the materials using the FTS. The Stycast 1090, Stycast 1266 and Stycast 2850FT have dielectric constants of 2.05, 2.60 and 4.95 respectively. We also successfully obtained epoxy with an intermediate dielectric constant by mixing two types of epoxy as shown in Figure \ref{fig:ar}. To get dielectric constant higher than 4.95, we mixed Stycast 2850FT with $\mathrm{SrTiO_3}$ filler which was shown to have a high dielectric constant \cite{Lee}. We were able to fabricate materials with a dielectric constant as high as 7.5. 

To test the multi-layer AR coating, we applied two layers of coating made from Stycast 1090 and Stycast 2850FT on flat 2 inch alumina sample which has a dielectric constant of 9.6. We then cooled the sample to 140~Kelvin using liquid nitrogen and measured transmission using the FTS. As shown in Figure \ref{fig:ar}, uncoated alumina shows high Fabry-Perot fringes due to high reflection whereas coated sample has high transmittance over a wide band. We are currently preparing 3-layer coatings in similar way to demonstrate even wider bandwidth.

To make a sufficiently precise coating on a lens, we designed a mold that leaves a thin gap between lens and mold. The cavity was made using a precision machined ball-ended mill which has 0.0005~inch tolerance. Combined with machining tolerance, we were able to create coatings with a tolerance of 0.001~inch, which is approximately $10\%$ of the thickness of the layer. We filled the cavity with the appropriate amount of epoxy measured by its weight. Then we inserted the lens in a mold that keeps its concentricity and accurate depth. Epoxy cures in a few hours in a $100^\circ \mathrm{C}$ oven. We repeated the process with molds with different spacings. 

To test its cryogenic adhesion property, we made twelve 2-layer coated lenses as shown in Figure \ref{fig:ar}. We kept one sample as a control, nine samples were slowly cooled to liquid nitrogen temperature in a 6~inch IR Labs dewar. Two additional samples were rapidly thermal cycled between room tempeature and liquid nitrogen temperature until failure. Failure for those samples occurred after 18 and 50 dunks. Nine samples went through ten slow thermal cycles in the dewar and all survived. These nine samples were then dunked 18 times in liquid nitrogen, bringing them back up to room temperature each time, and all survived. There was no change to the control sample that was kept at room temperature. We conclude that the adhesion of the coating is good enough for our applications given the results of this testing. 

\section{CONCLUSION}\label{sec:conclusion}
\begin{figure}
\begin{center}
\begin{tabular}{c}
\includegraphics[height=4cm,keepaspectratio]{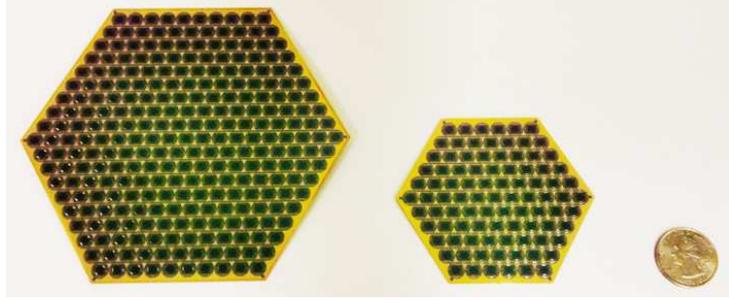}
\end{tabular}
\end{center}
\caption[example] 
{\label{fig:array} 
Photograph of the fabricated detector arrays and US 25 cent coin for size comparison. Larger hexagonal array is 13~cm across and it has 271 pixels. Smaller hexagonal array is 8cm across and it has 91 pixels. Both wafers has lumped diplexer (90~GHz and 150~GHz) with 1500~$\mu\mathrm{m}$ radius sinuous antenna.}
\end{figure} 
\begin{figure}
\begin{center}
\begin{tabular}{c}
\includegraphics[height=4cm,keepaspectratio]{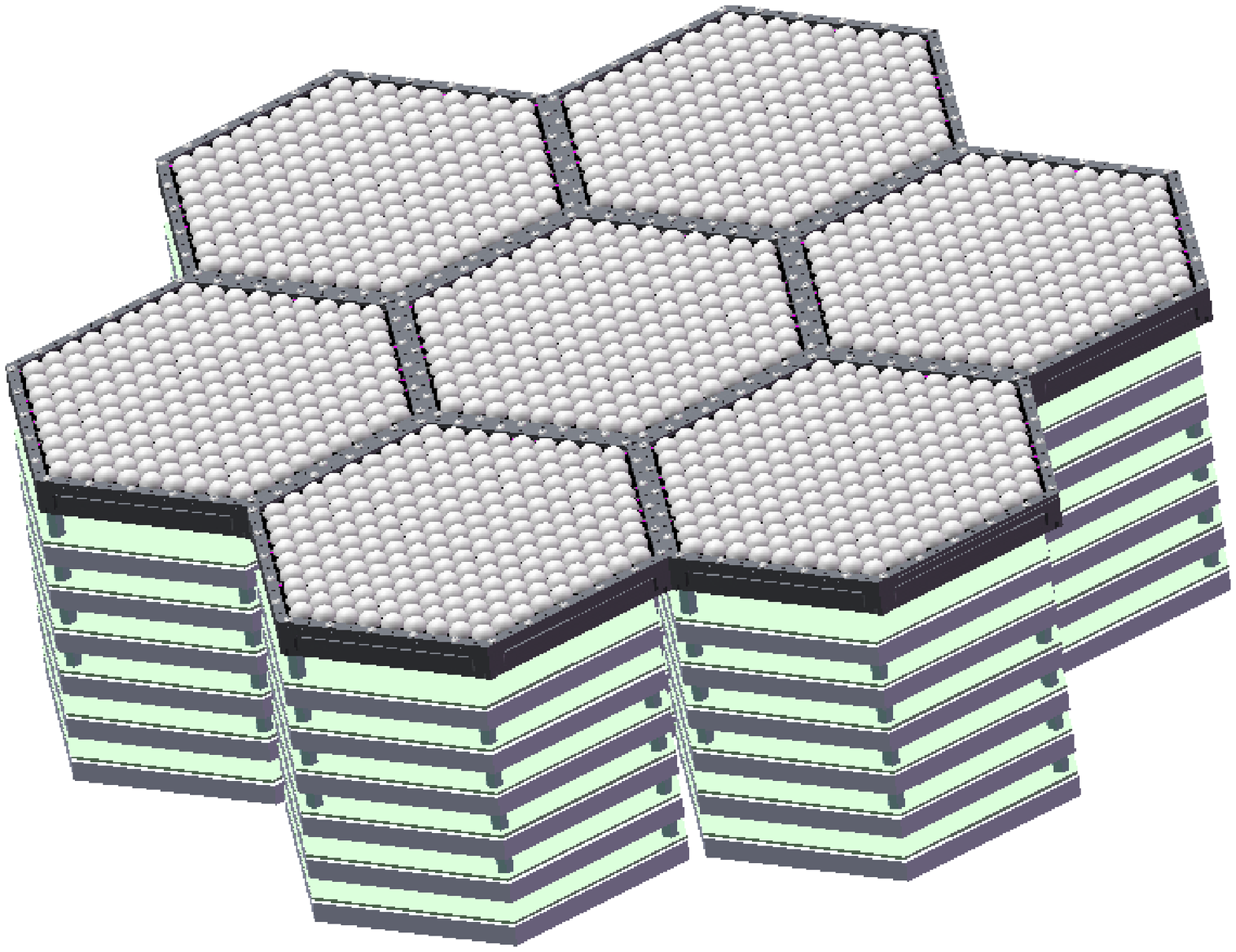}
\includegraphics[height=4cm,keepaspectratio]{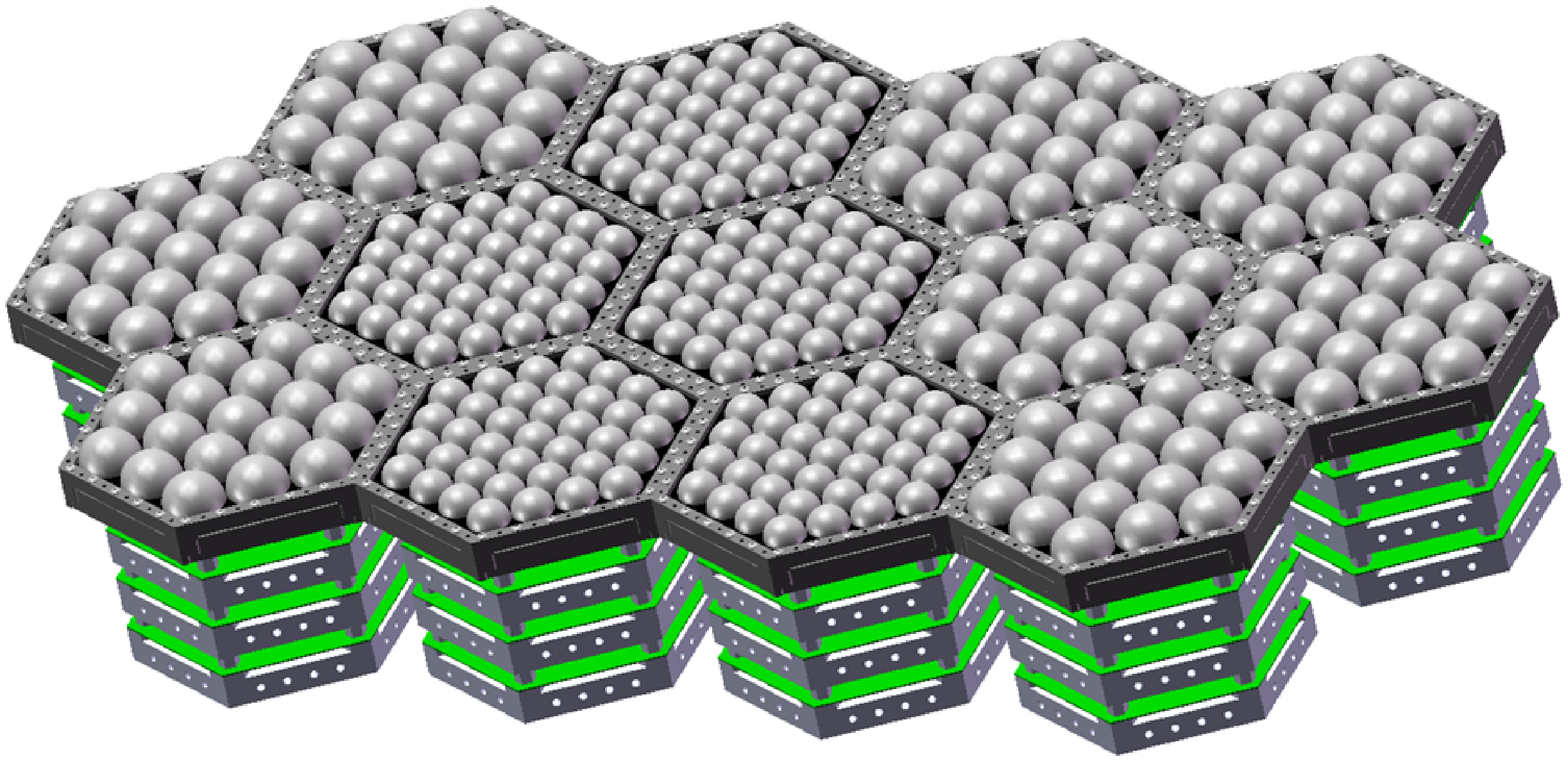}
\end{tabular}
\end{center}
\caption[example] 
{\label{fig:fp} 
CAD drawing of the proposed \pb-2 forcal plane (left) and \LB focal plane (right). \pb-2 focal plane mounts seven heagonal wafers with lumped diplexer with total of 7588 bolometers. \LB focal plane uses two types of detectors. There will be eight low frequency wafers with larger lenses and five high frequency wafers near the center of the focal plane. \LB focal plane will have 2022 bolometers total.}
\end{figure} 
To bring the proto-type multi-chroic pixels closer to readiness for future CMB experiments, we fabricated the arrays of multi-chroic detectors shown in Figure \ref{fig:array}. We fabricated these on 6-inch wafers. The larger array has a side-to-side length of 13cm, containing 271 pixels. The smaller array has a side-to-side length of 8cm, with 91 pixels. Each pixel uses a dual-polarized diplexer (90~GHz and 150~GHz bands) with a 1500~$\mu\mathrm{m}$ radius sinuous antenna. 
The next generation CMB B-mode experiments such as \Pb-2 and \LB are planning to use these multi-chroic detectors to increase their mapping speed. CAD drawings of the proposed focal planes for \Pb-2 and \LB are shown in Figure \ref{fig:fp}. \Pb-2 will use 7 of these 13~cm hexagonal arrays in its focal plane. The total number of bolometers will be 7588. \Pb-2 will achieve an array noise equivalent temperature (NET) of $ 6$~$\mu\mathrm{K}\sqrt{\mathrm{s}}$. \LB, a compact satellite CMB B-mode experiment, is planning to use two types of triplexer pixel: low frequency pixels (60,78 and 100~GHZ) and high frequency pixels (140, 190 and 280~GHz). The current design uses eight low frequency arrays and five high frequency arrays. There will be 2022 bolometers with an array NET of $ 1$~$\mu\mathrm{K}\sqrt{\mathrm{s}}$. 

We have successfully demonstrated multi-chroic detection with single pixels. With correct design of the antenna, we produced a detector with a round beam and low cross-pol performance. We have also presented a solution for broadband AR coatings, and we have fabricated arrays of multi-chroic detectors which will be used in future CMB experiments. 

\acknowledgments     
We acknowledge support from NASA through grant NNX10AC67G. Devices were fabricated at the Marvell Nanofabrication Laboratory at the University of California, Berkeley.

\bibliographystyle{spiebib}   
\bibliography{SPIE_Proceeding_Suzuki_V2}   

\begin{thebibliography}{10}

\bibitem{OBrientSPIE}
O'Brient, R., Ade, P., Arnold, K., Edwards, J., Engargiola, G., Holzapfel, W.,
  Lee, A.~T., Meng, X., Myers, M., Quealy, E., Rebeiz, G., Richards, P., and
  Suzuki, A., ``A dual-polarized multichroic antenna-coupled tes bolometer for
  terrestrial cmb polarimetry,'' {\em SPIE proceedings}  (2010).

\bibitem{Edwards}
Edwards, J., O'Brient, R., Lee, A., and Rebeiz, G., ``Dual polarized sinuous
  antennas on extended hemispherical silicon lenses,'' Antennas and
  Propagation, IEEE Transactions, submitted (2011).

\bibitem{Filipovic}
Filipovic, D., ``Double-slot antennas on extended hemispherical and elliptical
  silicon dielectric lenses,'' {\em Microwave Theory and Techniques, IEEE
  Transactions}~{\bf 41}(10),  1738--1749 (1993).

\bibitem{DuHamel}
Duhamel, R., ``Dual polarized sinuous antennas,'' (1987).
\newblock United States Patent 4658262.

\bibitem{Arnold}
Arnold, K., {\em Design and Deployment of the \Pb\ Cosmic Microwave Background
  Polarization Experiment}, PhD thesis, University of California, Berkeley
  (2010).

\bibitem{OBrient}
O'Brient, R., {\em A Log Periodic Focal-Plane Architecture for Cosmic Microwave
  Background Polarimetry}, PhD thesis, University of California, Berkeley
  (2010).

\bibitem{Kumar}
Kumar, S., Vayonakis, A., LeDuc, H., Day, P., Golwala, S., and Zmuidzinas, J.,
  ``Millimeter-wave lumped element superconducting bandpassfilters for
  multi-color imaging,'' {\em Applied Superconductivity, IEEE Transaction}~{\bf
  19}(3),  924--929 (2009).

\bibitem{Kerr}
Kerr, A., ``Surface impedance of superconductors and normal conductors in em
  simulators,'' tech. rep., ALMA (1999).
\newblock http://www.alma.nrao.edu/memos/.

\bibitem{Galbraith}
Galbraith, C. and Rebeiz, G., ``Higher order cochlea-like channelizing
  filters,'' {\em Microwave Theory and Techniques, IEEE Transactions}~{\bf
  56}(7),  1675--1683 (2008).

\bibitem{Lee}
Lee, S., Hyun, J., Kim, H., and Paik, K., ``A study on dielectric constants of
  $\mathrm{Epoxy/SrTiO_3}$ composite for embedded capacitor films (ecfs),''
  {\em Advanced Packaging, IEEE Transaction}~{\bf 30}(3),  428--433 (2007).

\end{thebibliography}

\end{document}